\documentclass[letter]{aa}

\usepackage{amsmath}
\usepackage[dvips]{graphicx}
\usepackage{natbib}
\bibpunct{(}{)}{;}{a}{}{,}
\usepackage[american]{babel}
\usepackage{txfonts}

\def\igr{IGR\,J17511$-$3057}

\def\XMM{{\em XMM-Newton}}

\def\chan{{\em Chandra}}

\def\rxte{{\em RXTE}}
\def\swift{{\em Swift}}

\begin{document}

\title{Swift monitoring of the new accreting millisecond X-ray pulsar IGR\,J17511$-$3057 in outburst}

\author{Bozzo, E. 
	\inst{1}
	\and
Ferrigno, C. 
\inst{1,2}
\and
Falanga, M. 
\inst{3}
\and
Campana, S. 
\inst{4}
\and 
Kennea, J. A. 
\inst{5}
\and 
Papitto, A. 
\inst{6,7} 
}

\authorrunning{E. Bozzo et al.}
\titlerunning{Swift observations of IGR\,J17511$-$3057 in outburst}
  \offprints{Enrico.Bozzo@unige.ch}

\institute{ISDC data centre for astrophysics, University of Geneva, 
	 chemin d'\'Ecogia, 16 1290 Versoix Switzerland
         \and
          IAAT, Abt.\ Astronomie, Universit\"at T\"ubingen
          Sand 1, 72076 T\"ubingen, Germany
          \and
          International Space Science Institute (ISSI) Hallerstrasse 6, CH-3012 Bern, Switzerland
          \and
          INAF - Osservatorio Astronomico di Brera, via Emilio Bianchi 46, I-23807 Merate (LC), Italy.
          \and 
          Department of Astronomy and Astrophysics, 525 Davey Lab, Pennsylvania State University, University Park, PA 16802, USA
          \and 
          Universit\'a degli Studi di Cagliari, Dipartimento di Fisica, SP Monserrato-Sestu, KM 0.7, 09042 Monserrato, Italy
          \and 
          INAF - Osservatorio Astronomico di Cagliari, Poggio dei Pini, Strada 54, 09012 Capoterra (CA), Italy
         }

\date{Received Oct. 21 2009; accepted Nov. 28 2009}

\abstract
	{A new accreting millisecond X-ray pulsar, IGR\,J17511$-$3057, was discovered in outburst on 2009 September 12 during 
  the \textsl{INTEGRAL} Galactic bulge monitoring programme.}
	{To study the evolution of the source X-ray flux and spectral properties during the outburst,  
	we requested a \swift\ monitoring of IGR\,J17511$-$3057.}
	{In this paper we report on the results of the first two weeks of monitoring the source.}
	{The persistent emission of IGR\,J17511$-$3057 during the outburst is modelled well with an absorbed blackbody 
	(kT$\sim$0.9~keV) and a power-law component ($\Gamma$$\sim$1-2), similar to what has been observed from other 
	previously known millisecond pulsars. \swift\ also detected three type-I Xray bursts from this source. 
	By assuming that the peak luminosity of these bursts is equal to the Eddington value for a pure helium 
	type-I X-ray burst, we derived an upper limit to the source distance of $\sim$10~kpc. 
	The theoretically expected recurrence time of the bursts according to the helium burst hypothesis 
	is 0.2-0.9~days, in agreement with the observations.}{} 
	
 \keywords{X-rays: binaries, pulsars: individual: IGR\,J17511$-$3057}

\maketitle

\section{Introduction}
\label{sec:intro} 

Accreting millisecond pulsars (AMSP) are neutron stars (NS) that accrete
mass from a low-mass ($\ll$1~$M_{\odot}$) companion star and that show
coherent pulsations at their millisecond spin period \citep{bhattacharya91}.
Since the discovery of the first AMSP in 1998, SAX~J1808.4$-$3658 \citep{wijnands98},  
11 other AMSPs were discovered \citep[see e.g.,][and Wijnands 2006 for reviews]{altamirano09,casella08,altamirano08}. 
All the AMSP are hosted in low-mass X-ray binaries 
(LMXBs) with typical orbital periods of either about 40~min or a few hours, and exhibit a transient X-ray 
emission with bright outbursts (10$^{36}$-10$^{37}$~erg/s) occurring on a time scale from two to more than ten years, and  
a typical duration of a few weeks \citep[see e.g.,][]{wijnands06}. However, longer activity episodes 
have also been recorded \citep[XTE\,J1807-294, see e.g.,][]{falanga05a}, and one  
source, HETE\,J1900.1-3455, has never returned in quiescence since its discovery 
in 2005 \citep{kaaret06,galloway07}. 
So far, only AMSP with orbital periods of a few hours have exhibited type-I X-ray 
bursts (the only exception being IGR~J00291+5934). This is believed to stem from the 
composition of the material that is accreted onto the NS that in turns depend on the 
nature of the donor star \citep[see e.g.,][and references therein]{galloway06}.   
The broad band X-ray (0.5-200 keV) spectra of AMSPs are
generally fit by a model consisting of one or two soft 
thermal components and a comptonized hard component.
The two soft components are associated with the radiation 
from the accretion disk and from the heated hot 
spots on the NS surface. The hard emission is likely to be produced
by thermal Comptonization of the soft photons emitted by the NS 
surface in the hot accretion column above the NS hot spots 
 \citep{gierlinski02,Poutanen03,gierlinski05,
falanga05a,falanga05b,falanga07,papitto09,patruno09,ibragimov09}.  

In this letter we report on the first $\sim$2 weeks of monitoring with \swift\ of the 12th 
newly discovered AMSP in outburst, IGR\,J17511$-$3057. 
We discuss the evolution of the source X-ray flux and spectrum from the onset of the outburst, 
when the source was discovered, up to the beginning of the outburst decay.
We also detected and analysed three type-I X-ray bursts.  

\subsection{The source IGR\,J17511$-$3057} 

\begin{table*}
\scriptsize
\caption{ \swift\ observation log of \igr.\ } 
\begin{tabular}{@{}lllllllllll@{}}
\hline
\noalign{\smallskip}
OBS ID & INSTR & START TIME & STOP TIME & EXP$^a$ & $N_{\rm H}$ & $\Gamma$ & $kT_{\rm BB}$ & $R_{\rm BB}$$^b$ & $F_{\rm obs}^{c}$ & $\chi^2_{\rm red}$/d.o.f. \\
      &       &            &           &  (sec)   &  (10$^{22}$~cm$^{-2}$) &  & keV & km & (erg/cm$^{2}$/s) & (C-stat/d.o.f.) \\
\noalign{\smallskip}
\hline
\noalign{\smallskip}
00031492001 & XRT/WT & 2009-09-13 19:52:24 & 2009-09-14 00:53:31 & 2.2E+03 & 0.6$\pm$0.1 & 1.3$^{+0.2}_{-0.3}$ & 1.0$\pm$0.1 & 4.6$\pm$0.8 & 4.9$^{+0.1}_{-0.4}$ & 1.2/447  \\
            & XRT/PC & 2009-09-13 21:29:48 & 2009-09-14 01:00:49 & 1.5E+03 & 1.0$\pm$0.2 & 1.5$\pm$0.1 & --- & --- & 4.8$\pm$0.3 & 0.9/81 \\

00031492002 & XRT/WT (1) & 2009-09-14 13:45:33 & 2009-09-14 13:49:50 & 2.5E+02 & 0.9$\pm$0.2 & 1.5$\pm$0.1 & --- & --- & 4.8$\pm$0.3 & 1.2/74\\

            & XRT/WT (2) & 2009-09-14 15:26:34 & 2009-09-14 15:29:01 & 1.5E+02 & 0.4$^{+0.3}_{-0.2}$ & 1.0$\pm$0.3 & --- &  --- & 4.5$\pm$0.8 & 1.0/19\\

            & XRT/PC & 2009-09-14 13:49:45 & 2009-09-14 15:37:21 & 1.6E+03 & 1.0$\pm$0.2 & 1.5$\pm$0.2 & --- & --- & 4.5$\pm$0.5 & 1.2/50 \\

00031492003 & XRT/WT (1) & 2009-09-15 15:31:43 & 2009-09-15 15:41:25 & 5.8E+02 & 0.4$\pm$0.2 & 0.9$^{+0.5}_{-1.4}$ & 0.9$\pm$0.1 & 6.0$^{+1.4}_{-1.8}$ & 4.8$^{+0.1}_{-2.3}$ & 1.1/174  \\

            & XRT/WT (2) & 2009-09-15 17:12:27 & 2009-09-15 17:26:00 & 6.1E+02 & 0.4$^{+0.2}_{-0.1}$ & 0.8$^{+0.6}_{-1.5}$ & 0.9$\pm$0.1 & 6.3$^{+0.9}_{-1.5}$ & 4.8$^{+0.1}_{-2.1}$ & 1.2/183  \\

            & XRT/PC & 2009-09-15 15:41:27 & 2009-09-15 15:49:56 & 5.0E+02 & 0.5$\pm$0.3 & 1.2$\pm$0.2 & --- & --- & 5.0$^{+0.4}_{-0.6}$ & 0.9/36 \\

00031492004 & XRT/WT (1) & 2009-09-16 17:16:29 & 2009-09-16 17:26:54 & 6.2E+02 & 0.3$\pm$0.2 & 0.5$^{+0.7}_{-2.2}$ & 0.9$\pm$0.1 & 5.5$^{+0.8}_{-1.8}$  & 4.5$^{+0.1}_{-2.2}$ & 1.1/170 \\

            & XRT/WT (2) & 2009-09-16 18:57:28 & 2009-09-16 19:12:00 & 8.7E+02 & 0.7$\pm$0.2 & 1.6$\pm$0.4 & 1.0$\pm$0.2 & 3.9$^{+1.0}_{-0.9}$  & 4.0$^{+0.2}_{-0.6}$ & 1.0/221 \\

           & XRT/PC & 2009-09-16 17:26:57 & 2009-09-16 17:34:28 & 4.2E+02 & 0.9$^{+0.4}_{-0.3}$ & 1.5$\pm$0.3 & --- & --- & 4.2$^{+0.2}_{-0.6}$ & 1.2/32 \\

00031492005 & XRT/WT (1) & 2009-09-18 20:21:26 & 2009-09-18 20:24:14 & 1.7E+02 & 0.9$\pm$0.2 & 1.4$\pm$0.2 & --- & ---  & 4.8$\pm$0.4 & 0.9/48 \\

           & XRT/WT (2) & 2009-09-18 23:34:25 & 2009-09-18 23:39:20 & 3.0E+02 & 0.9$^{+0.2}_{-0.1}$ & 1.5$\pm$0.1 & ---  & --- & 4.6$\pm$0.3 & 0.8/81 \\

           & XRT/PC & 2009-09-18 20:24:16 & 2009-09-18 23:43:57 & 1.1E+03 & 1.0$\pm$0.2 & 1.6$\pm$0.2 & --- & --- & 3.9$^{+0.2}_{-0.3}$ & 1.1/67 \\

00031492006 & XRT/WT & 2009-09-19 20:27:26 & 2009-09-19 22:25:50 & 1.4E+03 & 0.6$\pm$0.2 & 1.2$^{+0.3}_{-0.5}$ & 0.9$\pm$0.1 & 4.6$^{+1.4}_{-1.3}$ & 4.0$^{+0.1}_{-1.4}$ & 1.0/296 \\

           & XRT/PC & 2009-09-19 20:27:38 & 2009-09-19 22:26:57 & 3.0E+02 & 0.6$^{+0.5}_{-0.4}$ & 1.2$^{+0.4}_{-0.3}$ & --- & --- & 4.6$^{+0.4}_{-1.0}$ & 0.9/17 \\

00031492007 & XRT/WT (1) & 2009-09-20 09:18:13 & 2009-09-20 09:25:39 & 4.5E+02 & 0.8$^{+0.6}_{-0.3}$ & 1.9$^{+2.5}_{-0.8}$ & 1.1$\pm$0.2 & 3.6$^{+1.4}_{-0.8}$  & 3.6$^{+0.4}_{-1.7}$ & 0.9/110 \\

           & XRT/WT (2) & 2009-09-20 10:58:54 & 2009-09-20 11:01:28 & 1.5E+02 & 0.9$^{+0.3}_{-0.2}$ & 1.5$\pm$0.2 & ---  & --- & 4.0$^{+0.3}_{-0.6}$ & 1.2/38 \\

           & XRT/PC & 2009-09-20 09:25:42 & 2009-09-20 11:07:55 & 6.8E+02 & 0.9$\pm$0.2 & 1.5$\pm$0.2 & --- & --- & 3.6$^{+0.2}_{-0.4}$ & 1.1/47 \\

00031492008 & XRT/WT (1) & 2009-09-20 20:31:54 & 2009-09-20 20:34:14 & 2.0E+02 & 0.9$\pm$0.2 & 1.6$\pm$0.2 & --- & --- & 3.6$\pm$0.3 & 1.2/49 \\

           & XRT/WT (2) & 2009-09-20 22:08:54 & 2009-09-20 22:09:27 & 3.2E+01 & 1.0$^{+0.5}_{-0.4}$ & 1.9$\pm$0.5 & ---  & --- & 3.3$^{+0.5}_{-1.3}$ & (23.0/29) \\

           & XRT/PC & 2009-09-20 20:35:18 & 2009-09-20 22:14:57 & 5.2E+02 & 1.1$^{+0.4}_{-0.3}$ & 1.5$\pm$0.3 & --- & --- & 3.5$^{+0.3}_{-0.6}$ & 1.1/34 \\

00031492009 & XRT/WT (1) & 2009-09-21 06:10:54 & 2009-09-21 06:16:49 & 3.6E+02 & 0.9$\pm$0.2 & 1.6$\pm$0.2 & --- & --- & 3.7$\pm$0.2 & 1.3/87 \\

           & XRT/WT (2) & 2009-09-21 07:46:53 & 2009-09-21 07:52:37 & 3.4E+02 & 1.0$\pm$0.2 & 1.7$\pm$0.2 & ---  & --- & 3.5$\pm$0.2 & 1.1/80 \\

           & XRT/WT (3) & 2009-09-21 09:23:06 & 2009-09-21 11:04:42 & 3.9E+02 & 0.8$\pm$0.2 & 1.4$\pm$0.1 & ---  & --- & 3.9$\pm$0.2 & 1.1/93 \\

           & XRT/PC & 2009-09-21 06:16:50 & 2009-09-21 11:40:02 & 2.9E+03 & 1.1$\pm$0.2 & 1.7$\pm$0.1 & --- & --- & 3.2$\pm$0.2 & 0.9/113 \\
           
00031492010 & XRT/WT & 2009-09-28 10:16:53 & 2009-09-28 19:42:14 & 2.0E+02 & 1.1$\pm$0.3 & 1.8$\pm$0.3 & --- & --- & 2.3$\pm$0.3 & 1.1/31 \\

           & XRT/PC & 2009-09-28 10:17:20 & 2009-09-28 19:50:57 & 4.6E+03 & 1.0$\pm$0.2 & 1.5$\pm$0.1 & --- & --- & 2.4$^{+0.2}_{-0.1}$ & 0.9/101 \\

00031492011 & XRT/PC & 2009-09-30 18:23:35 & 2009-09-30 23:37:56 & 5.1E+03 & 1.1$\pm$0.1 & 1.7$\pm$0.1 & --- & --- & 2.0$\pm$0.1 & 1.2/204 \\

00371210000 & XRT/PC & 2009-09-30 18:35:26 & 2009-09-30 18:57:18 & 1.3E+03 & 1.1$\pm$0.2 & 1.7$\pm$0.2 & --- & --- & 2.2$\pm$0.1 & 0.9/70 \\
\noalign{\smallskip} 
\hline
\noalign{\smallskip} 
\multicolumn{11}{l}{{a}: the total exposure time of each observation; {b}: calculated 
by assuming a distance of 10\,kpc; {c}: the (absorbed) flux in the 0.5-10\,keV energy band in units of 10$^{-10}$.}
\end{tabular}
\label{tab:log}
\end{table*} 

IGR\,J17511$-$3057 was discovered on 2009 September 12 during 
the \textsl{INTEGRAL} Galactic bulge monitoring program \citep{baldovin09,kuulkers07}. 
The source was detected in both the \textsl{IBIS/ISGRI} and \textsl{JEM-X} mosaics, and its 3-100 keV energy spectrum was modelled  
by using a power law with index $2.0\pm0.2$. The corresponding flux was $1.1\times 10^{-9}\,\mathrm{erg\,s^{-1}\,cm^{-2}}$. 
A position within $2^\prime$ was determined, which allowed identification of IGR\,J17511$-$3057 as a new 
245\,Hz accreting pulsar by means of the \textsl{RXTE PCA} bulge scan data \citep{markwardt09}. 
Using this instrument, the source was initially detected on 2009 September 11, 
but not recognized as a new object due to the proximity of two previously known 
sources, XTE~J1751$-$305 and GRS~1747$-$312, the first 
of which is a 435\,Hz X-ray millisecond pulsar. 
The \textsl{RXTE} spectrum could be described by an absorbed power-law model with photon index 1.8 and 
a 2-10~keV flux of 4$\times$$10^{-10}\, \mathrm{erg\,s^{-1}\,cm^{-2}}$.  
A \textsl{Swift} ToO observation was carried out on 2009 September 13, and it allowed for a preliminary 
characterization of the soft (0.5-10~keV) X-ray spectrum 
and the discovery of the first thermonuclear type-I X-ray burst from the source \citep{bozzo09}. 
Burst oscillations at $\sim$245~Hz in other type-I X-ray bursts from 
this source were reported by \citet{watts09}. Further follow-up observations of IGR\,J17511-3057 were  
carried out later with \XMM\ \citep{papitto09b} and \rxte\ \citep{riggio09}. The latter provided 
the most precise ephemeris of the source and yielded a pulse frequency of 
244.83395157(7) Hz, an orbital period of 12487.5126(9)~s, and an $a \sin(i)$/c value of 275.194(3) lt-ms. 
The mass function of the system is thus 0.00107025(4)~$M_{\odot}$, giving a minimum companion 
mass of 0.13~$M_{\odot}$ (assuming an NS mass of 1.4~$M_{\odot}$ and errors at $1~\sigma$ c.l. in the last digit).    
The first accurate source position was determined through a \chan\ observation at  
$\alpha_{\rm J2000}$=17$^{\rm h}$51$^{\rm m}$08$\fs$66 and 
$\delta_{\rm J2000}$=-30${\degr}$57$\arcmin$41$\farcs$0 
\citep[1~$\sigma$ error of $\sim$0.6$\farcs$,][]{nowak09}, and allowed for identifying its  
infrared counterpart \citep{torres09}. Radio observations at this position did not result in any detection 
\citep{miller09}. 
\begin{figure}
\centering
\includegraphics[scale=0.32,angle=-90]{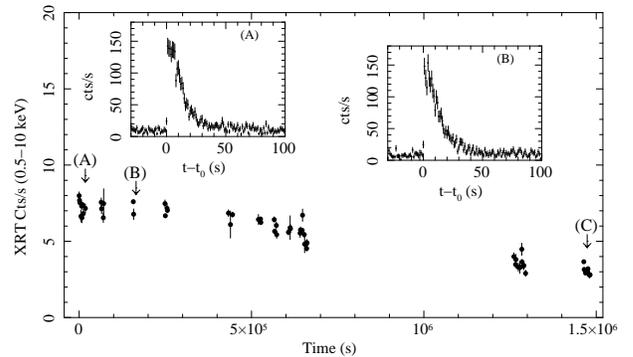}
\caption{\swift\,/XRT long-term light curve of the outburst of IGR\,J17511$-$3057 
(time bin 1000~s, start time 2009 Sept. 13 at 20:00:41 UTC). 
The arrows in the plot mark the time of the type-I X-ray bursts observed 
with \swift.\ The two inserts show a zoom of the first two type-I X-ray bursts 
(time bin is 1~s). The time on the X-axes of these inserts is 
measured from the $t_0$ of the bursts. These are 2009 Sept. 14 00:50:27 and 
2009 Sept. 15 at 17:17:19 (UTC), respectively. The third  
type-I X-ray burst (C) is reported in Fig.~\ref{fig:burstc}.}   
\label{fig:lcurve} 
\end{figure}

\section{Data analysis and results}
\label{sec:results} 

To monitor the X-ray flux of IGR\,J17511$-$3057 in outburst, we requested  
a 2~ks daily observation for the first week, and then two other observations of 
5~ks were scheduled during the second week when we noticed that the source X-ray flux 
was already decreasing after the beginning of the outburst (see Fig.~\ref{fig:lcurve}).   
A complete log of the observations is provided in Table~\ref{tab:log}.  
We processed all the \textsl{Swift} data by using standard procedures 
\citep{burrows05} and the latest calibration files available (caldb v. 20090407). 
The \textsl{Swift}/XRT data were analysed both in window-timing (WT) and photon-counting 
(PC) modes (processed with the {\sc xrtpipeline} v.0.12.3). 
We used \swift\,/BAT data accumulated only in {\sc event} mode, as the statistics of 
the data in {\sc survey} mode were too poor to provide any significant constraint on the 
source high-energy emission (15-150~keV).  

Filtering and screening criteria were applied by using {\sc ftools} (Heasoft v.6.6.3). 
We extracted source and background light curves and spectra by selecting event 
grades of 0-2 and 0-12,  respectively, for the WT and PC modes. 
Exposure maps were created through the {\sc xrtexpomap} task, and we used the latest 
spectral redistribution matrices in the {\sc heasarc} calibration database (v.011). 
Ancillary response files, accounting for different 
extraction regions, vignetting, and PSF corrections, were generated 
by using  the {\sc xrtmkarf} task.  
All PC data were affected by a strong pile-up, and corrected according 
to the technique developed by \citet{vaughan06}. 
We used the {\sc xrtlccorr} task to account for this 
correction in the background-subtracted light curves. 

\begin{figure}
\centering
\includegraphics[scale=0.32,angle=-90]{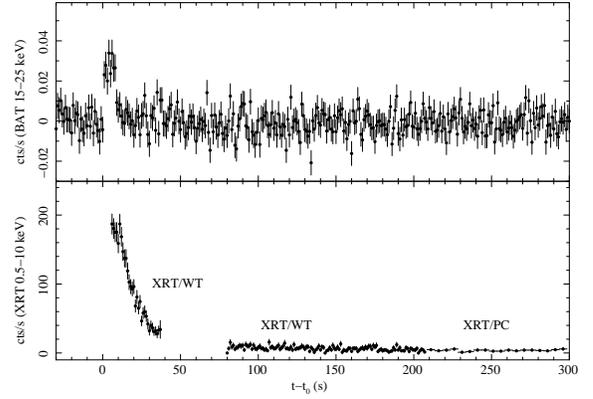}
\caption{The brightest type-I X-ray burst observed with \swift.\ 
The upper panel shows the BAT light curve (15-25~keV), 
whereas in the lower panel we reported the XRT light curve (0.5-10 keV).  
The time bin of the BAT and XRT/WT (XRT/PC) 
light curves is 1~s (5~s).
The start time of the burst is 2009 Sept. 30 18:31:57 (UTC).}   
\label{fig:burstc} 
\end{figure}

During the three type-I X-ray bursts we also checked a possible pile-up 
of the XRT/WT data, caused by the high source count rate. 
We extracted the source spectrum during the brightest part (6~s) of the three events  
by adopting a box-shaped extraction region in which we progressively excluded the 
first and then two, three, and four inner central pixels\footnote{see also http://www.swift.ac.uk/pileup.shtml}. 
These spectra were then fit with an absorbed black body model. We did not notice 
any significant pile-up in the spectral properties of the source at the peaks of the first two bursts. 
Only in the third burst (burst ``C'', see Fig.~\ref{fig:burstc}), we noticed that the XRT/WT data of the first 
two seconds of the burst were affected by a relatively strong pile-up. However, the time interval over which 
these data were accumulated is too short to apply the correction method described above and obtain a usable 
spectrum. Therefore, we discarded these first two seconds of observation. The peak luminosity 
we derive below for the third burst might thus have been somewhat higher than the value we reported, 
yielding a conservative upper limit on the distance we provide in Sect.~\ref{sec:discussion}. 
 
We performed a Fourier analysis of the persistent emission and of the type I X-ray burst emission of 
IGR\,J17511-3057 in order to search for the known 245 Hz periodicity of the source. 
We corrected the photons' arrival times for the orbital motion
of the source according to the solution given by \citet{riggio09}. 
During the type-I X-ray bursts, we could not detect any periodicity up to 283 Hz given a detection
threshold of 5~$\sigma$. This translates into an upper limit on any fractional amplitude of 
the pulsed signal of 0.23 (at 3~$\sigma$ confidence level), where the binning
effect of the temporal resolution of \swift\,/XRT in WT mode (1.8ms) has been
taken into account \citep{klis89}.
A similar analysis carried out on the persistent emission of the
source only led to a marginal 5~$\sigma$ detection of pulsations,
corresponding to an amplitude $\gtrsim$0.07 (3 sigma c.l.). We note that
the sensitivity to periodic signals is hampered here by the 
lower number of photons collected by \swift\,/XRT than by, e.g. RXTE/PCA, with 
which the pulsations and the burst oscillations 
were detected from this source (see Sect.~\ref{sec:intro}).    
In Fig.~\ref{fig:lcurve} we show the source light curve during about two  
weeks of monitoring.

\subsection{Persistent emission}
\label{sec:persistent}  

We extracted the source spectrum in the 0.5-10~keV energy band 
for each observation in Table~\ref{tab:log} excluding the time intervals in which a 
type I X-ray burst was detected. We excluded about $\sim$100~s before and 
after the bursts in the observations 00031492001 and 003. 
In those observations in which the XRT/WT data were accumulated 
during two or three different revolutions of the satellite, we extracted the X-ray spectrum 
separately to search for spectral variations. We indicated these   
different spectra in Table~\ref{tab:log} by using the notation ``1'', ``2'', and ``3''. 
The XRT/WT spectra with the higher statistics (observation ID: 00031492001, 003, 004, 006, 007) 
could not be fit by only using a simple absorbed power-law ($\Gamma$$\sim$1-2)  
model (reduced $\chi^2$$\sim$1.5-1.9, d.o.f. in the range 112-447), and the addition of a second component 
was required by the data. According to previous studies of AMSPs in outburst, we tried a model 
comprising an absorbed power law (PL) plus a black body ({\sc bbodyrad} in {\sc xspec}, 
hereafter BB) or a disk black body ({\sc diskbb} in {\sc xspec}) component. 
We found that the latter choice would need a highly improbable 
physical explanation to account for the large, inner accretion radius 
(a few hundred km) implied by the {\sc diskbb} model. 
As the spectra collected at higher statistics for this and similar sources 
show two soft components  
arising from the disk and the NS surface (see Sect.~\ref{sec:intro}), we interpret the soft component 
detected with \swift\ as originating in the NS hotspots. 
Our interpretation is strengthened as it provides an area 
for the BB-emitting region that is fully compatible with the NS surface. 
This is shown in Table~\ref{tab:log} (errors at 90\% c.l.). We checked that the fits to the \swift\,/XRT 
spectra cannot be improved significantly by introducing a Comptonization 
model instead of a simple power law 
above $\sim$2~keV. We tried a {\sc compTT} model in {\sc xspec} with different values of the 
seed photon temperature. The estimated column density 
is $\sim$0.5-1$\times$10$^{22}$~cm$^{-2}$, compatible with the Galactic absorption 
in the direction of the source \citep{dickey90}. The 0.5-10~keV X-ray flux of the 
source decreased slowly from 5.0$\times$10$^{-10}$~erg~cm$^{-2}$~s to 2.0$\times$10$^{-10}$~erg~cm$^{-2}$~s 
during the first two weeks of observation.

\subsection{Type I X-ray bursts}
\label{sec:bursts}  

\begin{table}
\centering
\scriptsize
\caption{The three type-I X-ray bursts parameters.}
\begin{tabular}{@{}lccc@{}}
\hline
\noalign{\smallskip}
Burst &  A & B & C \\
\hline
\noalign{\smallskip}
$\tau_\mathrm{lc}$ (s)$^a$    & $11.2 \pm0.8$ & $12.5\pm 0.8$ & $14.3\pm0.9$ \\
$F_{\rm peak}$ (10$^{-8}$ erg cm$^{-2}$ s$^{-1}$)$^b$  &  $3.0\pm0.6$  & $2.4\pm0.5$ & $3.1\pm 0.3$\\
$f_b$  (10$^{-7}$ erg cm$^{-2}$)$^c$ & $2.7\pm0.2$ &  $2.8\pm0.2$  & $4.8\pm0.3$\\
$F_\mathrm{pers}$ (10$^{-9}$ erg cm$^{-2}$ s$^{-1})$$^d$ & $3.1\pm0.4$  & $3.5\pm0.3$  & $0.95\pm0.09$  \\
$\gamma  \equiv F_{\rm pers}/F_{\rm peak}$  (10$^{-2}$) &  $10\pm3$ &  $14\pm4$ & $3.1\pm0.6$\\
$\tau$ (s) $\equiv f_b / F_{\rm peak} $  &  $9\pm3$ &   12$\pm3$ &   $16\pm3$\\
\noalign{\smallskip}
\hline
\noalign{\smallskip}
\multicolumn{4}{l}{$a$: Burst e-folding decay time. The $\chi^2_{\rm red}$/dof of the fits 
are 1.4/98, 1.2/86, and 1.3/29,} \\ 
\multicolumn{4}{l}{respectively, for the burst A,B, and C. $b$: Net unabsorbed peak flux, $c$: Net unabsorbed} \\ 
\multicolumn{4}{l}{burst fluence, and $d$: Unabsorbed persistent flux (all given in 0.1--100 keV).}\\ 
\end{tabular}
\label{tab:burst}
\end{table} 

We also detected in the \swift\ observations three type-I X-ray bursts.
The light curves of these bursts are reported in Fig.~\ref{fig:lcurve} and \ref{fig:burstc}.
The start time of the third burst was determined by \swift\,/BAT (XRT 
started observing the source only about $\sim$5~s later). 
We performed a time-resolved spectral analysis of the three bursts by accumulating the XRT/WT
spectra in intervals of different durations (from 1 to 10~s), depending on the source count rate. 
This time-resolved analysis did not reveal a clear signature of a photospheric radius 
expansion (PRE) in any of the three bursts \citep{lewin93}. 
The relevant parameters for each burst are reported 
in Table~\ref{tab:burst}. Here the peak flux of each burst was determined from a BB fit to the  
spectrum of the initial 6~s of each burst (fixing the $N_{\rm H}$ at 0.6, see Table~\ref{tab:log}). 
The persistent spectrum determined from the closest XRT/WT observation to each of 
the burst was used as a background in the fit. We indicated with $\tau_\mathrm{lc}$ 
the decay time of the burst measured by fitting the observed light curve 
with an exponential function, and $\tau$ is the duration of the burst 
\citep[see e.g.][]{lewin93}.  

\section{Discussion and conclusions}
\label{sec:discussion}

In this letter we reported on the results of the first $\sim$2 weeks of the \swift\ monitoring 
of the newly discovered AMSP IGR\,J17511$-$3057 in outburst. 
Regarding the persistent emission of this source, the higher statistics 
spectra measured with XRT/WT showed that two different components were required to fit the data. These comprise 
a BB emission that we interpreted as being produced onto the NS surface (as suggested by the size of 
the measured radius compatible with a hotspot origin), and a power-law component 
that is most likely caused by the comptonization of the soft emission in the NS accretion column 
(see Sect.~\ref{sec:intro}). This spectrum qualitatively agrees with observations of 
other AMSPs in outburst and with the preliminary results of the \XMM\ observation of 
IGR\,J17511$-$3057 reported by \citet{papitto09b}. 
At odds with their analysis, we could not detect the softest component of the spectrum that was modelled with   
a multicolor disk blackbody (temperature of 0.13$\pm$0.1~keV). This most likely stems from the lower statistics 
and the short exposure time of the XRT spectra compared to what is obtained with \XMM\ (exposure time $\sim$71~ks). 

From the analysis of the type-I X-ray bursts, we can estimate an upper limit on the source distance. 
In principle, determination of the distance can be obtained only when a burst 
undergoes a PRE; in this case, it is assumed that the bolometric
peak luminosity of the source is saturated at the Eddington limit, 
$L_{\rm Edd}$$\approx$3.8$\times$10$^{38}$$\mathrm{erg}{s}^{-1}$ \citep[as empirically derived by][]{kul03}. 
Unfortunately, in the case of IGR\,J17511$-$3057, our time-resolved analysis of the bursts could not detect any 
evidence of a PRE. 
From the measured orbital period (several hours) and the mass function 
of IGR\,J17511$-$3057, we can argue that the companion star in this system is most likely a 
hydrogen-rich brown dwarf, as suggested in the case of SAX\,J1808.4$-$3658 \citep{bildsten01}. 

Similar to the case for this source, we thus expect that the 
time elapsed between different bursts is long enough to allow the hot CNO burning 
to deplete the accreted hydrogen.
Therefore, the type I X-ray bursts of IGR\,J17511$-$3057
are most likely produced by the ignition of pure helium. 
Under the above assumptions, the peak luminosity of the 
brightest burst (C) can then be considered to be (at most) the Eddington 
value $L_{\rm Edd}$$\approx$3.8$\times$10$^{38}$$\mathrm{erg}^{-1}$, and 
the resulting upper limit on the source distance is $d$=10.1$\pm$0.5~kpc. 
For comparison, the theoretical value of this distance 
found by assuming an helium atmosphere and canonical NS parameters 
\citep[1.4 solar mass, and a radius of 10 km; see e.g.,][]{lewin93}, is
8.9$\pm$0.4~kpc. 
By using the above results, we can also evaluate the theoretically expected 
recurrence time of the bursts. With a distance of $d$=10~kpc,    
the persistent unabsorbed 0.1-100~keV flux of the 
source at the time of the bursts (A), (B), and (C)
would translate into a bolometric luminosity 
of $L_{\rm pers}\approx~3.7, 4.2, 1.1\times 10^{37}$~erg~s$^{-1}$, respectively. 

With these values at hand, we can estimate the local accretion rate per 
unit area, $\dot{m}$, through the relation $L_{\rm pers}=4\pi R^2\dot m(GM/R)/(1+z)$ 
\citep[with $z$=0.31 the NS gravitational redshift; see e.g.,][]{lewin93}.     
The ignition depths, $y_{\rm ign}$ for the three bursts can be calculated by using the equation 
$E_{\rm burst}=4\pi R^2y_{\rm ign}Q_{\rm nuc}/(1+z)$, where  
$E_{\rm burst}=4\pi d^2f_b$, $f_b$ is the fluence of the burst (see Table~\ref{tab:burst}), and   
$Q_{\rm nuc}\approx 1.6$ MeV corresponds to the nuclear
energy release per nucleon for complete burning of helium to iron
group elements \citep{wallace81}. We obtain $y_{\rm ign}$=2.2,2.4,4.0$\times$10$^8$~${\rm g\ cm^{-2}}$  
for the bursts (A), (B), and (C), respectively.  
For the above values of the local accretion rates and ignition depths, the expected 
recurrence time of the bursts is about   
$\Delta$t=($y_{\rm ign}$/$\dot m$)$(1+z)$$\simeq$0.2-0.9~days. 
This agrees with the recurrence time measured from the \XMM\ \citep[two bursts in $\sim$71~ks,][]{papitto09b} and 
\textsl{INTEGRAL} observations (4 bursts in $\sim$200\,ks carried out from 2009 September 16 to 2009 September 19, Falanga et al. 
2009, in preparation). 

Several more \swift\ observations of IGR\,J17511$-$3057 have already been planned.  
The results of this long-term monitoring campaign will be reported elsewhere 
(Campana et al. 2009, in preparation).

\section*{Acknowledgments}
We thank N. Gehrels and the \swift\ team for the rapid schedule of the observations of 
IGR\,J17511$-$3057, and an anonymous referee for useful comments. 
 CF has been supported by grant DLR~50~OG~0601.

\end{document}